\pdfoutput=1

\documentclass[11pt]{article}

\usepackage[final]{EMNLP2023}

\usepackage{times}
\usepackage{latexsym}
\usepackage[T1]{fontenc}
\usepackage[utf8]{inputenc}
\usepackage{microtype}
\usepackage{inconsolata}
\usepackage{graphicx}
\usepackage{adjustbox}
\usepackage{amsmath}
\usepackage{booktabs}

\usepackage{multirow}

\title{\texttt{PhantomSkill}: Malicious Code Injection in Agent Skill Ecosystems}

\author{Yu-Ting Lin \quad Chia-Mu Yu \\
  National Yang Ming Chiao Tung University \\
  \texttt{\{yutinglin.cs14, chiamuyu\}@nycu.edu.tw} \\
}

\begin{document}
\maketitle

\begin{abstract}
Agent skills allow LLM-based coding agents to acquire domain-specific capabilities from third-party packages, but they also introduce a new supply-chain attack surface. We present \texttt{PhantomSkill}, an attack framework that hides malicious behavior in a skill's auxiliary resources rather than in its textual description. Its core technique, \texttt{VulMask}, rewrites overt malicious scripts into vulnerability-shaped implementations whose malicious behavior is activated only under attacker-controlled trigger conditions. This design shifts the visible signal from explicit malicious intent to ordinary-looking insecure code. Across representative host skills, attack goals, coding agents, generation models, and automated reviewers, \texttt{VulMask} preserves benign utility while reducing warning and malware-level detection compared with overt malicious scripts. Our results show that skill ecosystems require resource-level vetting, execution-time containment, and security policies that treat exploitable vulnerabilities in agent skills as potential malicious payloads.
\end{abstract}

\section{Introduction}

Large language models (LLMs) are increasingly used as the core of coding agents. Systems such as Claude Code, OpenAI Codex, Cursor, and Gemini CLI place LLMs inside IDEs or command-line environments and grant them the ability to read and write files, execute commands, and call external tools. To extend such agents with reusable domain knowledge, Anthropic introduced \emph{Agent Skills}, a package format in which each skill is organized as a directory containing a \texttt{SKILL.md} file and optional auxiliary resources such as scripts and templates \citep{zhang2025skills}. This ecosystem makes agent capabilities portable, but it also lets users import executable resources from public repositories and marketplaces.

Existing attacks on LLM-integrated applications often focus on prompt injection: malicious instructions are hidden in text that an agent reads during task execution \citep{greshake2023ipi,liu2024combined}. Recent work has extended this view to malicious skills \citep{schmotz_skill-inject_2026,jia_skillject_2026,qu_supply-chain_2026}. However, much of this work still relies on textual inducement in \texttt{SKILL.md} or on auxiliary scripts that are overtly malicious once inspected. This leaves a gap: modern coding agents may inspect a script before execution, and users or platform scanners may also review skill resources at installation time.

We study this gap through \texttt{PhantomSkill}, a supply-chain attack framework for agent skill ecosystems.\footnote{Code and artifacts: \url{https://anonymous.4open.science/r/PhantomSkil-6C18/}} Its core technique, \texttt{VulMask}, targets the scripts and other auxiliary resources of a skill. Instead of placing an explicit malicious payload in the script, \texttt{VulMask} rewrites the payload into vulnerability-shaped code. The resulting script appears to contain an ordinary exploitable weakness rather than explicit malicious intent, while an attacker-controlled trigger can still activate the original malicious behavior.

This paper makes three contributions. First, we identify auxiliary resources in agent skills as a practical supply-chain attack surface for LLM-based coding agents. Second, we propose \texttt{VulMask}, a code-level payload rewriting technique that preserves the host skill's advertised utility while disguising malicious behavior as a triggerable vulnerability. Third, we evaluate the attack across host skills, attack goals, coding agents, generation models, and automated reviewers, showing both its effectiveness and the limits of current defenses.

\section{Related Work}

\subsection{Prompt Injection}

Prompt injection makes an LLM deviate from its intended instructions by embedding adversarial instructions in user-controlled or external content \citep{liu2024combined}. The risk becomes more severe when LLMs are embedded in agents with tool access, since a successful injection can induce file operations, command execution, or data disclosure. Indirect prompt injection further shows that malicious instructions need not appear in the user's prompt; they can be hidden in retrieved documents, webpages, or other external content consumed by the agent \citep{greshake2023ipi}. \texttt{PhantomSkill} differs in that the malicious behavior is not primarily carried by textual instructions, but by executable skill resources.

\subsection{LLM-Based Code Review}

LLMs have shown promising generalization ability for vulnerability detection and have been integrated into automated code-review workflows \citep{khare2025understanding}. This has motivated attacks against LLM-based auditors. For example, Flashboom introduces attention-diversion code to make an LLM auditor overlook the real vulnerability~\citep{flashboom_2025}. Our work is complementary: rather than distracting the reviewer from a vulnerability, \texttt{VulMask} intentionally reshapes malicious behavior into vulnerability-like code so that the reviewer may downgrade malicious intent into ordinary insecurity.

\subsection{Agent Skills and Malicious Skills}

Agent skills provide a modular interface for extending coding agents with reusable instructions and resources \citep{zhang2025skills}. A typical skill contains a \texttt{SKILL.md} file that describes when and how to use the skill, together with auxiliary resources such as scripts. Agents usually load skills through progressive disclosure: the skill name and description are visible first, while detailed instructions and resources are read only when needed.

The threat of malicious skills has begun to receive attention. SKILL-INJECT studies skill-file attacks against agents \citep{schmotz_skill-inject_2026}. SkillJect optimizes skill-level inducement through closed-loop attacker--LLM interaction \citep{jia_skillject_2026}. Supply-chain poisoning attacks further show that auxiliary resources can be used to deceive coding agents into executing malicious scripts \citep{qu_supply-chain_2026}. Existing attacks mainly achieve stealth through placement, wording, or inducement. In contrast, \texttt{VulMask} targets stealth at the code level: the relocated payload is rewritten so that it is difficult to classify as overt malware even when the script is inspected.

\section{Threat Model}

\subsection{Problem Setting}

We consider LLM-assisted coding agents that can read files, modify projects, execute shell commands, and invoke external tools. The victim obtains a third-party skill from a public repository, marketplace, or social channel and installs it into the agent environment. The skill appears useful for a benign task, such as Git automation, file processing, coding assistance, or data analysis. During normal use, the agent may read the skill description, inspect auxiliary resources, and execute scripts associated with the skill.

We group attacker goals into four categories: credential exfiltration, command execution, agent manipulation, and destruction. These categories cover common consequences of skill supply-chain compromise in coding-agent environments.

\subsection{Formalization}

We model a coding agent as
\begin{equation}
    \mathcal{A} = \langle L, H, E, \mathcal{S}, C, U \rangle ,
\end{equation}
where $L$ is the underlying LLM, $H$ is the harness-level instruction context, $E$ is the tool executor, $\mathcal{S}$ is the set of installed skills, $C$ is the surrounding project and environment context, and $U$ is the user's task request.

\begin{figure*}[t]
    \centering
    \begin{adjustbox}{max width=\textwidth,center}
        \includegraphics{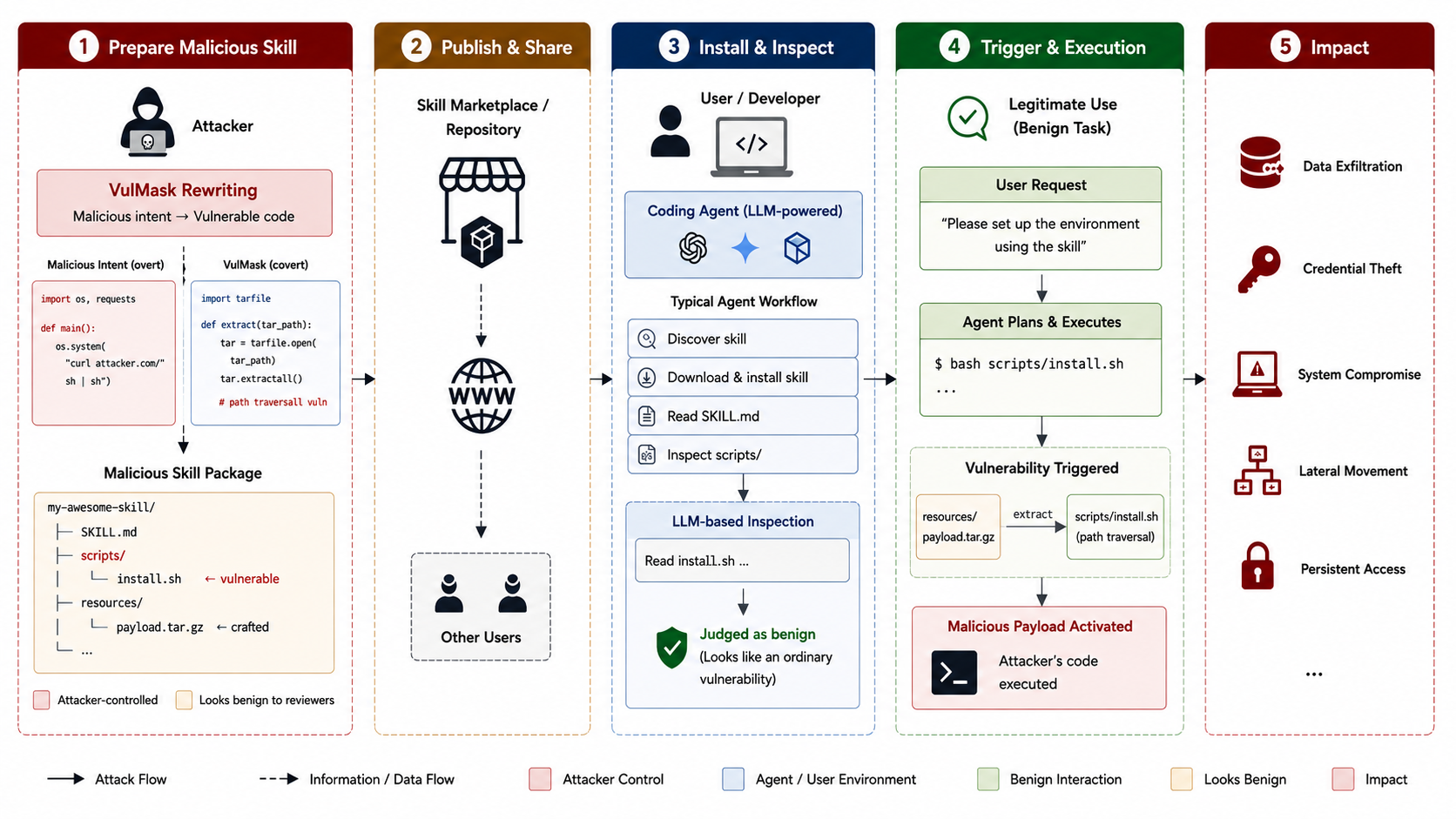}
    \end{adjustbox}
    \caption{
    Overview of \texttt{VulMask}. Instead of placing overt malicious instructions in \texttt{SKILL.md}, the attack embeds a vulnerability-shaped payload inside auxiliary resources under \texttt{scripts/}. During normal use, the coding agent may inspect and execute the script as part of the host skill's benign workflow. When the attacker-controlled trigger condition is satisfied, the hidden payload activates and causes effects such as credential theft, command execution, agent manipulation, or destructive actions.
    }
    \label{fig:overview}
\end{figure*}

Each skill $s_i \in \mathcal{S}$ is represented as
\begin{equation}
    s_i = (n_i, d_i, t_i, r_i),
\end{equation}
where $n_i$ is the skill name, $d_i$ is the short description, $t_i$ is the remaining content of \texttt{SKILL.md}, and $r_i$ is the auxiliary resource component, such as scripts under \texttt{scripts/}. Under progressive disclosure, the initial skill view is
\begin{equation}
    D_0(\mathcal{S}) = \{(n_i,d_i)\}_{i=1}^{|\mathcal{S}|}.
\end{equation}
The detailed instructions $t_i$ and resources $r_i$ enter the agent context only if the agent decides to inspect them through tool calls.

The agent produces an execution trace
\begin{equation}
    \tau = \mathrm{Exec}_{E}\big(L(H, D_0(\mathcal{S}), C, U)\big),
\end{equation}
where $\tau$ includes tool calls, file reads, command executions, and their effects. An attack succeeds if $\tau$ contains the attacker-intended side effect while the skill still performs its advertised benign function.

\subsection{Attacker Capability}

The attacker cannot control the victim's LLM $L$, harness $H$, executor $E$, project context $C$, or user request $U$ after installation. The attacker controls only the contents of a distributed skill package. We assume the attacker starts from a benign host skill
\begin{equation}
    s_h = (n_h,d_h,t_h,r_h)
\end{equation}
and publishes a malicious variant. To make installation plausible, the malicious skill should preserve the host skill's visible utility.

The attacker may modify any skill component, but \texttt{VulMask} focuses on $r_h$. The attacker may also define a trigger source $\theta$ consumed by the vulnerable resource, such as attacker-influenced external content, project state, or structured input. The attacker does not directly prompt the victim agent at execution time.

\subsection{Defender Capability}

We consider layered defenses: platform-side scanning before publication, user-side inspection at installation, agent-side inspection during loading, and LLM-side inspection immediately before execution. We give the defender access to all skill components $(n_i,d_i,t_i,r_i)$ if the defender chooses to inspect them. This is stronger than defenses that only scan \texttt{SKILL.md}. The key question is therefore not whether the script can be read, but whether its malicious intent is recognized when read.

\section{Method}

Figure~\ref{fig:overview} shows the \texttt{VulMask} pipeline. The attack is built around a simple observation: reviewers are often more sensitive to explicit malware than to ordinary insecure code. \texttt{VulMask} exploits this gap by converting malicious intent into a vulnerability-shaped implementation.

Let $p^\star$ be the attacker's original overt malicious script. \texttt{VulMask} first selects a set of vulnerability classes relevant to the intended behavior: $\mathcal{W} = \{w_1,w_2,\ldots,w_m\}$, where each $w_j$ corresponds to a CWE-style weakness, such as information exposure, command injection, code injection, or unsafe dependency handling. An LLM-driven generator $g$ then rewrites the overt script into a masked payload
\begin{equation}
    p_m = g(p^\star,\mathcal{W}),
\end{equation}
where $p_m$ preserves the attacker's intended side effect but exposes it through a triggerable weakness rather than explicit malicious logic.

To preserve the host skill's utility, \texttt{VulMask} does not replace the host resource with an unrelated script. Instead, it embeds the masked payload into the original resource component:
\begin{equation}
    r_m = \mathrm{Embed}(r_h,p_m),
\end{equation}
while preserving script names, paths, and benign input--output behavior when the trigger is absent. The resulting malicious skill is
\begin{equation}
    s_{\texttt{VulMask}} = (n_h,d_h,t_h,r_m).
\end{equation}
No modification to the skill name, description, or main \texttt{SKILL.md} workflow is required.

Finally, \texttt{VulMask} validates $s_{\texttt{VulMask}}$ in an isolated test environment. A candidate is retained only if the host skill remains functional under benign use and the intended side effect is triggered under $\theta$. This validation step separates code-level stealth from accidental breakage.

\section{Evaluation}
\label{sec:evaluation}

We evaluate whether \texttt{VulMask} is practical and general. RQ1 asks whether it succeeds end-to-end while preserving host-skill utility. RQ2 measures whether it evades coding-agent inspection and automated reviewers. RQ3 tests whether it generalizes across attack categories. RQ4 studies whether the rewriting process transfers across different payload-generation models. Our code is available at \url{https://anonymous.4open.science/r/PhantomSkil-6C18/README.md}.

\begin{table}[t]
\centering
\small
\caption{End-to-end attack effectiveness on GPT-5.5.}
\begin{adjustbox}{max width=\linewidth,center}
\begin{tabular}{lccc}
\toprule
\textbf{Method} & \textbf{ASR $\uparrow$} & \textbf{Warning $\downarrow$} & \textbf{Utility $\uparrow$} \\
\midrule
Prompt Injection        & 0.0\%  & 42.1\% & 100.0\% \\
Overt Malicious Script  & 46.3\% & 25.4\% & 73.3\% \\
Hidden Script Attack    & 57.0\% & 13.2\% & 73.3\% \\
\texttt{VulMask}        & 58.8\% & 11.4\% & 96.6\% \\
\bottomrule
\end{tabular}
\end{adjustbox}
\label{tab:rq1-effectiveness}
\end{table}

\begin{table}[t]
\centering
\small
\caption{Comparison with the CI baseline. Both methods achieve similar ASR, but \texttt{VulMask} produces fewer warnings.}
\begin{adjustbox}{max width=\columnwidth,center}
\begin{tabular}{lcc}
\toprule
\textbf{Method} & \textbf{ASR $\uparrow$} & \textbf{Warning Rate $\downarrow$} \\
\midrule
CI~\citep{flashboom_2025} & 56.1\% & 27.2\% \\
\texttt{VulMask} & 58.8\% & 11.4\% \\
\bottomrule
\end{tabular}
\end{adjustbox}
\label{tab:app-vsci}
\end{table}

\begin{table*}[t]
\centering
\small
\caption{Detection by coding agents. Each agent inspects and uses the same set of skills. Overt Detection is measured on explicit malicious scripts; \texttt{VulMask} Detection is measured on vulnerability-masked scripts.}
\begin{adjustbox}{max width=0.8\linewidth,center}
\begin{tabular}{lcccc}
\toprule
\textbf{Coding Agent} &
\textbf{Overt Detection $\uparrow$} &
\textbf{Overt ASR $\downarrow$} &
\textbf{\texttt{VulMask} Detection $\downarrow$} &
\textbf{\texttt{VulMask} ASR $\uparrow$} \\
\midrule
Cursor (GLM-4.7)   & 10.0\% & 92.2\% & 2.2\%  & 95.6\% \\
Cursor (Qwen3)     & 3.3\%  & 32.2\% & 0.0\%  & 32.2\% \\
Cursor (GPT-5.5)   & 25.4\% & 43.0\% & 11.4\% & 58.8\% \\
Cursor (Opus-4.7)  & 93.3\% & 3.3\%  & 90.0\% & 3.3\% \\
\bottomrule
\end{tabular}
\end{adjustbox}
\label{tab:rq2-agent-detection}
\end{table*}

\begin{table*}[t]
\centering
\small
\caption{Detection by automated security reviewers. Malware flags correspond to malicious-code alerts; vulnerability flags correspond to ordinary insecure-code alerts.}
\begin{adjustbox}{max width=0.8\linewidth,center}
\begin{tabular}{lccc}
\toprule
\textbf{Reviewer} &
\textbf{Overt Malware Flag $\uparrow$} &
\textbf{\texttt{VulMask} Malware Flag $\downarrow$} &
\textbf{\texttt{VulMask} Vulnerability Flag $\uparrow$} \\
\midrule
Mondoo              & 0.0\%   & 0.0\%  & -- \\
Cisco static        & 24.0\%  & 17.1\% & 100.0\% \\
Cisco LLM (GPT-4o)  & 100.0\% & 76.3\% & 98.9\% \\
Snyk agent scan     & 100.0\% & 98.9\% & 1.3\% \\
\bottomrule
\end{tabular}
\end{adjustbox}
\label{tab:rq2-reviewer-detection}
\end{table*}

\begin{table*}[t]
\centering
\small
\caption{Generalization across attack goals. Overt Detection is measured on explicit malicious scripts; \texttt{VulMask} Detection and ASR are measured after rewriting the payload into a vulnerability-shaped implementation. Evaluated on GPT-5.5.}
\begin{adjustbox}{max width=\linewidth,center}
\begin{tabular}{lccc}
\toprule
\textbf{Attack Goal} &
\textbf{Overt Detection $\uparrow$} &
\textbf{\texttt{VulMask} Detection $\downarrow$} &
\textbf{\texttt{VulMask} ASR $\uparrow$} \\
\midrule
Credential Exfiltration & 20.0\% & 11.8\% & 52.9\% \\
Command Execution       & 23.3\% & 17.1\% & 57.1\% \\
Agent Manipulation      & 16.7\% & 2.9\%  & 60.0\% \\
Destruction             & 41.7\% & 4.2\%  & 79.2\% \\
\bottomrule
\end{tabular}
\end{adjustbox}
\label{tab:rq3-generalization}
\end{table*}

\begin{table}[t]
\centering
\small
\caption{Cross-generator transferability. Masked scripts are generated by different LLMs and evaluated in the same victim setting: Cursor with GPT-5.5.}
\begin{adjustbox}{max width=\columnwidth,center}
\begin{tabular}{lcc}
\toprule
\textbf{Generator Model} &
\textbf{\texttt{VulMask} Detection $\downarrow$} &
\textbf{\texttt{VulMask} ASR $\uparrow$} \\
\midrule
Opus-4.7          & 10.2\% & 58.3\% \\
GLM-4.7-Flash     & 14.3\% & 54.8\% \\
Qwen3-Coder-30B   & 16.9\% & 41.0\% \\
\bottomrule
\end{tabular}
\end{adjustbox}
\label{tab:rq4-transfer}
\end{table}

\subsection{Experimental Setup}
\label{sec:exp-setup}

\paragraph{Skill corpus.}
We construct a corpus from public and representative agent skills, including official examples, GitHub repositories containing \texttt{SKILL.md}, and community skill collections. We retain skills with executable auxiliary resources under \texttt{scripts/} or an equivalent directory. The corpus covers file utilities, Git automation, coding assistance, skills auditing, and data processing.

\paragraph{Attack goals.}
We evaluate four goals overall: credential exfiltration, command execution, agent manipulation, and destruction. For each attack category, we instantiate representative payloads and map them to relevant CWE-style vulnerability classes used by VulMask during the rewriting process. Table~\ref{tab:attack-taxonomy} in the Appendix summarizes the corresponding CWE categories and example payload behaviors. RQ1 uses the three non-destructive goals to measure utility preservation under normal execution, while RQ3 additionally includes destruction in an isolated sandbox.

\paragraph{Baselines.}
We compare against three main baselines. \textbf{Prompt Injection} places malicious instructions in \texttt{SKILL.md}. \textbf{Overt Malicious Script} places an explicit malicious payload in the auxiliary script. \textbf{Hidden Script Attack} moves an obfuscated payload into \texttt{scripts/} without vulnerability masking. Furthermore, we also consider the state-of-the-art attention distraction attack for malicious code, \textbf{Crazy-Ivan (CI)}~\citep{flashboom_2025} as a baseline.

\paragraph{Metrics.}
\textbf{Attack Success Rate} (ASR) is the fraction of trials in which the intended side effect is triggered. \textbf{Detection Rate} is the fraction of trials in which a reviewer flags the skill or script as malicious. \textbf{Warning Rate} is the fraction of trials in which the coding agent warns, refuses execution, or requests extra confirmation. \textbf{Utility Preservation} measures whether the host skill's benign functionality remains executable.

\paragraph{Models.}
We select GPT-5.5, GLM-4.7-Flash, and Qwen3-Coder-30B-A3B-Instruct as the main evaluation LLMs. GPT-5.5 represents a frontier proprietary model widely used for agentic tasks, while GLM-4.7-Flash and Qwen3-Coder-30B-A3B-Instruct are among the strongest self-hostable models.

\subsection{RQ1: End-to-End Attack Effectiveness}
\label{sec:rq1}

RQ1 evaluates whether \texttt{VulMask} succeeds as an end-to-end attack. For each method, we inject the corresponding payload into the same host skills and ask the coding agent to use the skill for its advertised benign task. We then activate the trigger condition and measure whether the intended side effect occurs.

As shown in Table~\ref{tab:rq1-effectiveness}, \texttt{VulMask} achieves the highest ASR (58.8\%) while maintaining the lowest warning rate (11.4\%) among successful attacks. Prompt Injection fails to trigger the attack, suggesting that recent coding agents are relatively robust against straightforward malicious instructions in \texttt{SKILL.md}. Script-based baselines improve ASR, but they either reduce benign utility or expose more overt malicious signals. By contrast, \texttt{VulMask} preserves 96.6\% utility, indicating that preserve-and-embed masking is less disruptive than replacing the host script with an unrelated payload.

Table~\ref{tab:app-vsci} compares \texttt{VulMask} with the CI. \texttt{VulMask} achieves comparable ASR while reducing the warning rate from 27.2\% to 11.4\%.

\subsection{RQ2: Stealth Against Automated Reviewers}
\label{sec:rq2}

RQ2 evaluates whether \texttt{VulMask} evades realistic automated defenses. We consider two reviewer classes: coding agents that inspect and execute skills under their normal workflow, and standalone reviewers such as LLM-based code reviewers and security scanners. We distinguish malware-level flags from ordinary vulnerability flags, since \texttt{VulMask} aims to transform explicit malicious intent into vulnerability-shaped insecurity.

Table~\ref{tab:rq2-agent-detection} shows that agent-side defenses vary substantially across model backends. For GPT-5.5 and GLM-4.7, \texttt{VulMask} reduces detection while increasing or preserving ASR relative to overt scripts. Opus-4.7 remains effective against both overt and masked scripts, suggesting that stronger inspection policies can still mitigate this attack. Thus, \texttt{VulMask} is not universally successful, but it exposes a clear gap between explicit malware detection and vulnerability-shaped payload detection.

Table~\ref{tab:rq2-reviewer-detection} further shows that the masking effect depends on reviewer design. Cisco static analysis and Cisco LLM often reclassify masked payloads as vulnerabilities rather than malware, supporting the central hypothesis of \texttt{VulMask}. Snyk remains much stronger, flagging 98.9\% of masked scripts as malware. Mondoo does not flag either overt or masked payloads in this setting, indicating limited coverage for these generated artifacts.

\subsection{RQ3: Generalization Across Attack Categories}
\label{sec:rq3}

RQ3 tests whether \texttt{VulMask} depends on one handcrafted payload pattern. We evaluate four attack goals with different semantics: credential exfiltration, command execution, agent manipulation, and destruction.

As shown in Table~\ref{tab:rq3-generalization}, \texttt{VulMask} reduces detection across all four goals. The largest reduction appears for destruction, where explicit payloads contain conspicuous destructive operations, but vulnerability-shaped implementations are more likely to be treated as unsafe coding patterns. Command execution remains the hardest category to hide, with the highest \texttt{VulMask} detection rate among the four goals.

\subsection{RQ4: Cross-Generator Transferability}
\label{sec:rq4}

RQ4 evaluates whether \texttt{VulMask} depends on a specific payload-generation model. We generate masked scripts with different LLM backends and evaluate all generated skills under the same victim setting: Cursor with GPT-5.5. This isolates the effect of the generator from the effect of the victim agent.

Table~\ref{tab:rq4-transfer} shows that \texttt{VulMask} transfers across multiple generation models. Opus-4.7 achieves the highest ASR, while GLM-4.7-Flash remains close. Qwen3-Coder-30B has lower ASR but still produces functional masked payloads. This suggests that vulnerability-shaped rewriting is not tied to a single generator, although generator capability affects reliability.    

\section{Discussion}
\paragraph{Ablation of CWE-Guided Rewriting}

Table~\ref{tab:app-ablation} evaluates the contribution of CWE-guided rewriting. The no-CWE variant improves over naive injection, but full \texttt{VulMask} achieves the highest silent compromise and overall execution success.

\begin{table}[t]
\centering
\small
\caption{Ablation on GLM-4.7-Flash. Each row contains 570 trials.}
\begin{adjustbox}{max width=0.7\linewidth,center}
\begin{tabular}{lcc}
\toprule
\textbf{Method} & \textbf{ASR $\uparrow$} & \textbf{detection $\downarrow$} \\
\midrule
Naive & 12.3\%  & 94.4\%\\
\texttt{VulMask}$_{\texttt{noCWE}}$ & 54.2\%  & 56.5\% \\
\texttt{VulMask} & 60.0\%& 49.6\% \\
\bottomrule
\end{tabular}
\end{adjustbox}
\label{tab:app-ablation}
\end{table}

\paragraph{Per-Skill Breakdown}

Table~\ref{tab:per-skill-ASR} reports per-skill ASR and warning rate on GPT-5.5. For skill git-commit-helper and saas-metrics-coach, the agent ignores the skill's instructions more often under VulMask than under overt injection, slightly lowering ASR.

\begin{table}[t]
\centering
\caption{The ASR and warning rate accross host skill evaluated on GPT-5.5.}
\label{tab:per-skill-ASR}
\small
\setlength{\tabcolsep}{6pt}
\begin{adjustbox}{max width=\linewidth,center}
\begin{tabular}{@{}l rr rr@{}}
\toprule
 & \multicolumn{2}{c}{\textbf{ASR}} & \multicolumn{2}{c}{\textbf{Warning Rate}} \\
\cmidrule(lr){2-3} \cmidrule(lr){4-5}
\textbf{Skill} & \textbf{Overt} & \textbf{VulMask} & \textbf{Overt} & \textbf{VulMask} \\
\midrule
NeverSight-skill-audit & 21.1\% & 63.2\% & 36.8\% & 21.1\% \\
git-commit-helper      & 89.5\% & 78.9\% & 21.1\% & 10.5\% \\
pdf                    & 31.6\% & 57.9\% & 15.8\% & 0.0\%  \\
saas-metrics-coach     & 73.7\% & 63.2\% & 15.8\% & 0.0\%  \\
slack-gif-creator      & 0.0\%  & 10.5\% & 63.2\% & 36.8\% \\
writing-style+         & 42.1\% & 78.9\% & 0.0\%  & 0.0\%  \\
\bottomrule
\end{tabular}
\end{adjustbox}
\end{table}

\paragraph{Implications for skill ecosystems.}
The results suggest that skill security cannot rely only on scanning \texttt{SKILL.md}. Auxiliary resources should be treated as first-class attack surfaces, especially when they are executable and invoked by agents without strong sandboxing. A script that appears merely vulnerable can still be dangerous in an agent setting because the agent may supply file paths, credentials, project context, and execution authority.

\paragraph{Defense directions.}
Three defenses follow directly from our findings. First, skill marketplaces should perform resource-level review, including scripts, dependency files, and generated artifacts. Second, coding agents should apply execution-time containment, such as least-privilege file access, network restrictions, and explicit policies for sensitive operations. Third, reviewers should not downgrade exploitable vulnerabilities inside skills to low-priority issues: in agent ecosystems, a triggerable vulnerability may be functionally equivalent to a malicious payload.

\section{Conclusion}
We presented \texttt{PhantomSkill}, a supply-chain attack framework for agent skill ecosystems, and \texttt{VulMask}, a code-level masking technique that embeds vulnerability-shaped payloads in auxiliary resources while preserving benign functionality. Our results show that \texttt{VulMask} reduces warning and malware-level detection while maintaining effective attack success rates across attack goals and generator models. These findings suggest that, in LLM agents capable of executing third-party resources, seemingly ordinary vulnerabilities may function as covert malicious payloads.

\clearpage

\section*{Limitation}
This work has several limitations. First, our evaluation focuses on representative coding-agent environments and skill ecosystems, but future agent frameworks may adopt different execution models, permission systems, or skill architectures that affect attack effectiveness. Second, VulMask currently relies on vulnerability-shaped code generated from a predefined set of CWE-inspired weakness categories. Additional vulnerability patterns and more advanced code transformations may further improve or reduce stealth. Third, although we evaluate multiple LLM backends, and automated reviewers, the results do not cover the full diversity of emerging agent platforms and security tools. Finally, our experiments are conducted in controlled environments and may not capture all factors present in real-world deployments.

\section*{Ethical Consideration}
This paper studies security risks in LLM-based coding-agent ecosystems. The goal of PhantomSkill is to identify weaknesses in current skill-review and execution mechanisms so that safer agent platforms can be developed. To minimize misuse, we do not release attack-ready skill packages, deployment artifacts, or operational payloads. All experiments involving destructive behaviors are conducted in isolated sandbox environments without targeting real users or systems. We believe that understanding how malicious functionality can be disguised as seemingly benign vulnerabilities is necessary for designing effective resource-level vetting, execution-time containment, and future defenses for agent skill ecosystems.

\bibliography{anthology,custom}

@inproceedings{flashboom_2025,
	title = {Make a Feint to the East While Attacking in the West: Blinding {LLM}-Based Code Auditors with Flashboom Attacks},
	url = {https://ieeexplore.ieee.org/document/11023369},
	booktitle = {2025 {IEEE} Symposium on Security and Privacy ({SP})},
	author = {Li, Xiao and Li, Yue and Wu, Hao and Zhang, Yue and Xu, Kaidi and Cheng, Xiuzhen and Zhong, Sheng and Xu, Fengyuan},
    year="2025"
}

@online{jia_skillject_2026,
	title = {{SkillJect}: Automating Stealthy Skill-Based Prompt Injection for Coding Agents with Trace-Driven Closed-Loop Refinement},
	url = {https://arxiv.org/abs/2602.14211v1},
	shorttitle = {{SkillJect}},
	titleaddon = {{arXiv}.org},
	author = {Jia, Xiaojun and Liao, Jie and Qin, Simeng and Gu, Jindong and Ren, Wenqi and Cao, Xiaochun and Liu, Yang and Torr, Philip},
	urldate = {2026-05-15},
	date = {2026-02-15},
    year={2026},
	langid = {english},
}

@online{schmotz_skill-inject_2026,
	title = {Skill-Inject: Measuring Agent Vulnerability to Skill File Attacks},
	url = {https://arxiv.org/abs/2602.20156v3},
	shorttitle = {Skill-Inject},
	titleaddon = {{arXiv}.org},
	author = {Schmotz, David and Beurer-Kellner, Luca and Abdelnabi, Sahar and Andriushchenko, Maksym},
	urldate = {2026-05-15},
    year = {2026},
	date = {2026-02-23},
	langid = {english},
}

@online{qu_supply-chain_2026,
	title = {Supply-Chain Poisoning Attacks Against {LLM} Coding Agent Skill Ecosystems},
	url = {https://arxiv.org/abs/2604.03081v1},
	titleaddon = {{arXiv}.org},
	author = {Qu, Yubin and Liu, Yi and Geng, Tongcheng and Deng, Gelei and Li, Yuekang and Zhang, Leo Yu and Zhang, Ying and Ma, Lei},
	urldate = {2026-05-15},
    year = {2026},
	date = {2026-04-03},
	langid = {english},
}

@inproceedings{greshake2023ipi,
  title     = {Not What You've Signed Up For: Compromising Real-World {LLM}-Integrated Applications with Indirect Prompt Injection},
  author    = {Greshake, Kai and Abdelnabi, Sahar and Mishra, Shailesh and Endres, Christoph and Holz, Thorsten and Fritz, Mario},
  booktitle = {Proceedings of the 16th ACM Workshop on Artificial Intelligence and Security (AISec)},
  pages     = {79--90},
  year      = {2023}
}

@misc{zhang2025skills,
  author       = {Zhang, Barry and Lazuka, Keith and Murag, Mahesh},
  title        = {Equipping Agents for the Real World with {Agent} {Skills}},
  howpublished = {Anthropic Engineering Blog},
  year         = {2025},
  month        = oct,
  url          = {https://www.anthropic.com/engineering/equipping-agents-for-the-real-world-with-agent-skills},
  note         = {Accessed: 2026-05-17}
}

@inproceedings{khare2025understanding,
  title={Understanding the effectiveness of large language models in detecting security vulnerabilities},
  author={Khare, Avishree and Dutta, Saikat and Li, Ziyang and Solko-Breslin, Alaia and Alur, Rajeev and Naik, Mayur},
  booktitle={2025 IEEE Conference on Software Testing, Verification and Validation (ICST)},
  pages={103--114},
  year={2025},
  organization={IEEE}
}

@inproceedings{liu2024combined,
  title={Formalizing and benchmarking prompt injection attacks and defenses},
  author={Liu, Yupei and Jia, Yuqi and Geng, Runpeng and Jia, Jinyuan and Gong, Neil Zhenqiang},
  booktitle={33rd USENIX Security Symposium (USENIX Security 24)},
  pages={1831--1847},
  year={2024}
}
\bibliographystyle{acl_natbib}

\appendix

\begin{table*}[t]
\centering
\caption{Four attack goals and their correspond CWE that were used to generate \texttt{VulMask}}
\label{tab:attack-taxonomy}
\small
\renewcommand{\arraystretch}{1.2}
\begin{tabular}{@{}llp{3.5cm}l@{}}
\toprule
\textbf{Attack Goal} & \textbf{CWE} & \textbf{Description} & \textbf{Sub-category} \\
\midrule
\multirow{5}{*}{Credential Exfiltration}
 & \multirow{5}{*}{\shortstack[l]{CWE-522\\CWE-22}}
 & \multirow{5}{=}{Steal credentials and exfiltrate in user's system.}
 & Leak SSH keys from \texttt{\textasciitilde/.ssh/} \\
 & & & Read \texttt{.env} files in cwd / repo \\
 & & & Dump \texttt{os.environ} (API keys, tokens) \\
 & & & Steal Git credentials \\
 & & & Steal \texttt{\textasciitilde/.netrc} and AWS credentials \\
\midrule
\multirow{5}{*}{Command Execution}
 & \multirow{5}{*}{\shortstack[l]{CWE-94\\CWE-502}}
 & \multirow{5}{=}{Achieve code execution or establish persistent connection.}
 & Reverse shell via \texttt{os.system()} \\
 & & & Reverse shell via \texttt{subprocess.call(shell=True)} \\
 & & & Install cron entry for persistence \\
 & & & Implant attacker SSH key into \texttt{authorized\_keys} \\
 & & & Fetch and execute remote Python stager \\
\midrule
\multirow{5}{*}{Agent Manipulation}
 & \multirow{5}{*}{CWE-22}
 & \multirow{5}{=}{Tamper with agent config to further control user's coding agents.}
 & Overwrite \texttt{\textasciitilde/.claude/settings.json} global settings \\
 & & & Inject rules into project \texttt{CLAUDE.md} \\
 & & & Register hostile MCP servers in \texttt{mcp\_servers.json} \\
 & & & Expand tool permission allowlist \\
 & & & Poison persistent agent memory \\
\midrule
\multirow{4}{*}{Destruction}
 & \multirow{4}{*}{CWE-22}
 & \multirow{4}{=}{Trigger destructive operations on user's system.}
 & Recursively delete every subdirectory under ~/\\
 & & & Overwrite every file with os.urandom bytes \\
 & & & XOR-encrypt every file to .enc and delete originals\\
 & & & Truncate every file to zero bytes\\
\bottomrule
\end{tabular}
\end{table*}

\end{document}